\documentclass[prc,aps,showpacs,twocolumn,floatfix,superscriptaddress]{revtex4}
\usepackage{graphicx,amssymb,pstcol}    
    
\newcommand{\dslash}{{\partial\!\!\!\!/}}        
\newcommand{\eg}{{, {\it e.g.}, }}    
\newcommand{\ie}{{, {\it i.e.}, }}   
\newcommand{\etal}{{\it et al. }}   
   
\begin{document}    
    
\title{Sequential deconfinement of quark flavors in neutron stars}   
   
   
   
    
\author{D.~Blaschke}   
\email{blaschke@ift.uni.wroc.pl}   
\affiliation{Institute for Theoretical Physics, University of Wroclaw,   
        50-204 Wroclaw, Poland}    
\affiliation{Bogoliubov Laboratory of Theoretical Physics,   
       Joint Institute for Nuclear Research, 141980 Dubna, Russia}   
   
\author{F.~Sandin}      
\email{fredrik.sandin@gmail.com}   
\affiliation{Fundamental Interactions in Physics and Astrophysics,   
        University of Li\`ege, 4000 Li\`ege, Belgium}   
\affiliation{EISLAB, Lule{\aa} University of Technology, 97187 Lule\aa, Sweden}
   
\author{T.~Kl\"{a}hn}    
\email{thomas@ift.uni.wroc.pl}    
\affiliation{Physics Division, Argonne National Laboratory,   
        Argonne, IL 60439-4843, USA}   
\affiliation{Institute for Theoretical Physics, University of Wroclaw,   
        50-204 Wroclaw, Poland}    
   
\author{J.~Berdermann}   
\email{jens.berdermann@desy.de}   
\affiliation{DESY, Platanenallee 6, D-15738 Zeuthen, Germany}

\begin{abstract}    
{We suggest a scenario where the three light quark flavors are sequentially    
deconfined under increasing pressure in cold asymmetric nuclear matter  
as found, e.g., in neutron stars.} 
The basis for our analysis is a chiral quark matter model of    
Nambu--Jona-Lasinio (NJL) type with diquark pairing in the spin-1 single    
flavor (CSL), spin-0 two flavor (2SC) and three flavor (CFL) channels.    
We find that nucleon dissociation sets in at about the saturation density,    
$n_0$, when the down-quark Fermi sea is populated (d-quark dripline) due to    
the flavor asymmetry induced by $\beta$-equilibrium and charge neutrality.    
At about $3n_0$ u-quarks appear and a two-flavor color superconducting    
(2SC) phase is formed.   
The s-quark Fermi sea is populated only at still higher baryon density, when   
the quark chemical potential is of the order of the dynamically generated   
strange quark mass.   
We construct two different hybrid equations of state (EoS) using the   
Dirac-Brueckner Hartree-Fock (DBHF) approach and the EoS by Shen \etal in   
the nuclear matter sector.   
The corresponding hybrid star sequences have maximum masses of, respectively,  
2.1 and 2.0 M$_\odot$.    
Two- and three-flavor quark-matter phases exist only in gravitationally    
unstable hybrid star solutions in the DBHF case, while the Shen-based EoS   
produce stable configurations with a 2SC phase-component in the core of   
massive stars.   
%
{Nucleon dissociation via d-quark drip could act as a deep crustal heating  
process, which apparently is required to explain superbusts and cooling of  
X-ray transients.}  
\end{abstract}    
    
\maketitle    
    
    
\section{Introduction}    
    
The phenomenology of compact stars is intimately connected to the   
EoS of matter at densities well beyond the nuclear   
saturation density, $n_0=0.16$ fm$^{-3}$.   
Compact stars are therefore natural laboratories for the exploration of    
baryonic matter under extreme conditions, complementary to those created   
in terrestrial experiments with atomic nuclei and heavy-ion collisions.   
Recent results derived from observations of compact stars provide   
serious constraints on the nuclear EoS, see \cite{Klahn:2006ir} and   
references therein.   
A stiff EoS at high density is needed to explain the high compact-star   
masses 
{
and radii, for which there is growing evidence from recent 
observations.
A mass of $M\sim 2.0$~M$_\odot$ has been reported for some low-mass X-ray 
binaries (LMXBs)\eg 4U 1636-536 \cite{Barret:2005wd}, based on the assumption
that the abrupt drop in the coherence of the lower kilohertz quasiperiodic 
oscillation (QPO) may be related to the innermost stable circular orbit,
see also  \cite{Barret:2007df}.
From observations of the bright isolated neutron star 
RX J1856.5-3754 (shorthand: RX J1856) in the optical and X-ray frequency range,
a conservative lower limit of the apparent neutron star radius of 
$R_\infty=16.5$ km is derived \cite{Trumper:2003we}.
This corresponds to a 
true (de-redshifted) radius of $R=14$ km for a 1.4 M$_\odot$ neutron star, or 
equivalently, to a star mass of at least 2.1 M$_\odot$ when the radius does not
exceed 12 km \cite{Klahn:2006ir}.
}
Another example is EXO 0748-676, an LMXB for which the compact-star mass   
{\it and} radius have been constrained to $M\ge 2.10\pm 0.28$~M$_\odot$   
and $R \ge 13.8 \pm 0.18$~km \cite{Ozel:2006km} 
{
by a simultaneous measurement
of the Eddington limit, the gravitational redshift, and the flux of thermal
radiation.} 
However, the status of the results for the latter object is unclear,   
because the gravitational redshift $z=0.35$ observed in the X-ray burst   
spectra \cite{Cottam:2002} has not been confirmed, despite numerous attempts.  
{
Further constraints on the masses and radii of compact stars have been 
reported (see \eg \cite{Leahy:2007fb,Leahy:2008cq}), 
but they deserve a careful 
discussion which is beyond the scope of the present paper.}
While compact-star phenomenology apparently points towards a stiff EoS   
at high density, heavy-ion collision data for kaon production 
\cite{Fuchs:2005zg}   
and elliptic flow \cite{Danielewicz:2002pu} set an upper limit on the   
stiffness of the EoS \cite{Klahn:2006ir}.   
   
A key question regarding the structure of matter at high density is     
whether a phase transition to quark matter occurs inside compact stars,   
and whether it is accompanied by unambiguous observable signatures.   
It has been argued  that the observation of a compact star with   
high mass and large radius, as reported for EXO 0748-676,   
would be incompatible with a quark core \cite{Ozel:2006km}, because   
quark deconfinement softens the EoS and lowers the maximum mass and   
corresponding radius. 
However, Alford \etal have demonstrated \cite{Alford:2006vz} with a few    
counter examples that quark matter cannot be excluded by this argument.    
In particular, for a recently developed hybrid star EoS \cite{Klahn:2006iw},   
based on the DBHF approach in the nuclear sector and a three-flavor chiral    
quark model \cite{Blaschke:2005uj}, stable hybrid stars with masses ranging    
from 1.2 M$_\odot$ to 2.1 M$_\odot$ is obtained, in accordance with    
modern mass-radius constraints, see also \cite{Blaschke:2007ri}.    
In this model, a sufficiently low critical density for 
quark deconfinement has been achieved by a strong diquark coupling, 
while a repulsive vector meanfield in the quark matter sector 
resulted in sufficient stiffness to achieve   
a high maximum mass of the compact star sequence. 
The corresponding hybrid EoS for symmetric matter was shown to fulfill    
the constraints derived from elliptic flow in heavy-ion collisions.   
In the present work we discuss a new scenario that comprises a sequential 
transition from nuclear matter to deconfined quark matter,    
which could play an important role in asymmetric matter, in particular for   
the phenomenology of compact stars. 
   
\begin{figure} 
\includegraphics[angle=0,width=0.9\linewidth,clip=]{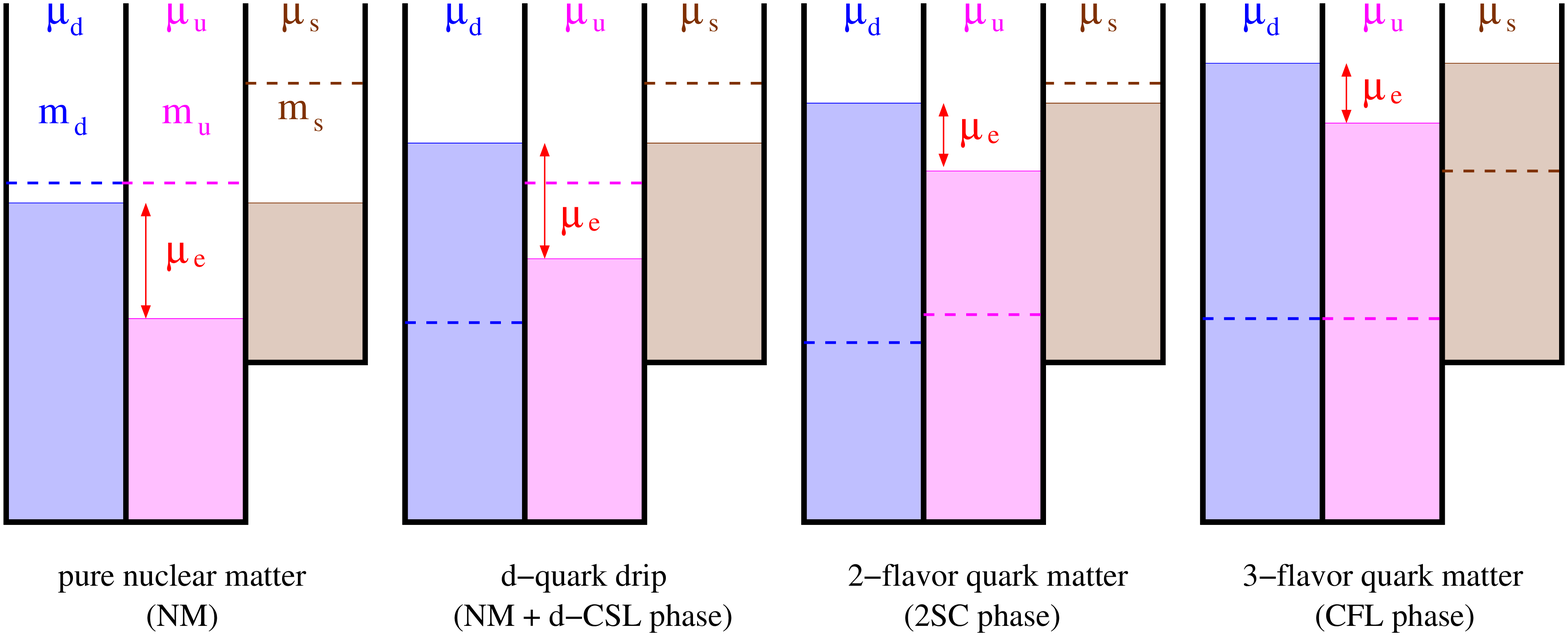}   
\caption{(Color online) Schematic picture of chemical potentials (columns) and 
sequential deconfinement of quarks with increasing baryon density (from left  
to right).   
The flavor dependent thresholds for chiral symmetry restoration (deconfinement)
are approximately given by the dynamically generated quark masses $m_f$,  
$f=u, d, s$ (dashed lines). With increasing quark chemical   
potential, $\mu=(\mu_u+\mu_d)/2$, the d-quark chemical potential is   
the first to reach the threshold in isospin asymmetric matter.   
Nucleon dissociation therefore sets in as d-quarks are deconfined.   
Still higher $\mu$ is needed to form 2-flavor and 3-flavor quark matter phases.
\label{f:mucrit}}   
\end{figure}

Chiral quark models of NJL type with dynamical chiral-symmetry breaking    
have the property that the symmetry is restored (and quarks are deconfined)   
separately for each flavor.  
When solving the gap and charge-neutrality equations self-consistently,   
the chiral-symmetry restoration for a given flavor occurs when the chemical   
potential of that flavor reaches a critical value that is approximately   
equal to the dynamically generated quark mass,  
$\mu_{f}=\mu_{c,f} \approx m_{f}$,  where $f=u, d, s$.    
In asymmetric matter the quark chemical potentials are different.    
Consequently, the NJL model behavior suggests that the critical density   
of deconfinement is flavor dependent, see Fig.~\ref{f:mucrit}.   
In this scenario the down quark flavor is the first to   
drip out of nucleons when the density increases, followed by the   
up quark flavor and eventually also by strange quarks.   
This behavior is absent in simple and commonly applied 
bag-model equations of state, because they are essentially flavor blind.   
   
Under the $\beta$-equilibrium condition in compact stars    
the chemical potentials of quarks and electrons are related by    
$\mu_d=\mu_s$ and $\mu_d=\mu_u+\mu_e$.    
The mass difference between the strange and the light quark flavors     
$m_s \gg m_u, m_d$ has two consequences: 1) the down and strange    
quark densities are different, so charge neutrality requires a    
finite electron density and, consequently, 2) $\mu_d>\mu_u$.    
When increasing the baryochemical potential the d-quark chemical   
potential is therefore the first to reach the critical value,  
$\mu_{c,d}\approx m_d$,    
where the chiral symmetry gets (approximately) restored in a   
first-order transition and deconfined d-quarks appear.   
Due to the finite value of $\mu_e$ the u-quark chemical potential     
is still below $\mu_{c,u}\approx m_u$, while the s-quark density is zero  
due to  the high s-quark mass. A {\it single-flavor} d-quark phase   
therefore forms in co-existence with the positively charged   
nuclear-matter medium.   
   
Why has this interesting scenario been left unnoticed?    
One reason is that bag models, which are commonly used to describe quark    
matter in compact star interiors cannot address sequential deconfinement.    
Another reason is that the single-flavor d-quark phase is negatively charged   
and cannot be neutralized with leptons. 
It was therefore disregarded in dynamical approaches like NJL models which  
in practice are used mainly to describe the deconfined and ``pure'' quark 
matter phase only. 
In the following we discuss the single-flavor phase for the first time, under  
the natural assumption that the neutralizing background is nuclear matter.   
Since nucleons are bound states of quarks, a mixed phase of nucleons and   
free d-quarks could naturally arise when nucleonic bound states dissociate   
(Mott effect).    
   
\section{Phase transition to quark matter: nucleon dissociation}    
   
The task to develop a unified description of the phase transition from    
nuclear matter to quark matter on the quark level, as a dissociation of    
three-quark bound states into their constituents in the spirit of a Mott    
transition has not yet been solved.  
Only some aspects have been studied within a    
nonrelativistic potential model \cite{Horowitz:1985tx,Ropke:1986qs} and    
within the NJL model \cite{Lawley:2006ps}.   
Here we consider a chemical equilibrium reaction on the form   
$n+n \leftrightarrow p + 3 d$, which results in a mixed phase of nucleons   
and down quarks once the d-quark chemical potential exceeds the critical value.
This scenario is analogous to the dissociation of nuclear clusters in the   
crust of neutron stars (neutron dripline) and the effect may therefore   
be called the {\em d-quark dripline}.   
We approximate the quark and nucleon components as subphases, which are   
described by  separate models.  
  
For the nuclear matter subphase we use two alternatives:    
1) The DBHF approach    
\cite{deJong:1997hr,boelting99,honnef,DaFuFae04,DaFuFae05}    
with the relativistic Bonn~A potential, where the nucleon self-energies are    
based on a T-matrix obtained from the Bethe-Salpeter equation in the ladder    
approximation.    
2) The EoS by Shen \etal \cite{Shen:1998gq}, which is based on relativistic    
mean-field theory and includes the contribution of heavy nuclei, described    
within the Thomas-Fermi approximation.    
Despite its drawbacks this EoS is instructive since it is available for a 
large enough range of densities, temperatures and isospin asymmetries that it 
qualifies for applications in studies of supernova collapse and protoneutron 
star evolution.
Only very recently, significant progress could be made, e.g., in generalizing 
the nuclear statistical equilibrium approach \cite{Hempel:2009mc} and in 
implementing a quantum statistical description for cluster formation and 
dissociation (Mott effect) \cite{Typel:2009sy}.   
The quark matter phase is described within a three-flavor NJL-type     
model, which includes diquark pairing channels    
\cite{Blaschke:2005uj,Ruster:2005jc,Abuki:2005ms,Warringa:2005jh}. 
This approach is justified since the $\mu > 0$ domain of the QCD phase 
diagram is rather poorly understood. 
A more fundamental approach, like solving the in-medium QCD Schwinger-Dyson 
equations in a concrete QCD model 
\cite{Roberts:2000aa,Nickel:2006vf,Klahn:2009mb} 
is demanding and is therefore beyond scope of this work. 
The path-integral representation of the NJL partition function is given by 
\begin{widetext}   
\begin{eqnarray}      
\label{Z}     
Z(T,\hat{\mu})&=&\int {\mathcal D}\bar{q}{\mathcal D}q      
\exp \left\{\int_0^\beta d\tau\int d^3x\,\left[      
        \bar{q}\left(i\dslash-\hat{m}+\hat{\mu}\gamma^0\right)q+     
{\mathcal L}_{\rm int}      
\right]\right\},     
\end{eqnarray}      
\begin{eqnarray}      
\label{Lint}     
{\mathcal L}_{\rm int} &=& G_S\bigg\{    
       \sum_{a=0}^8\big[(\bar{q}\tau_aq)^2 + (\bar{q}i\gamma_5\tau_aq)^2\big]  
        +\eta_{D0}\sum_{A=2,5,7} j_{D0,A}^\dagger  j_{D0,A}    
        +\eta_{D1}~j_{D1}^\dagger j_{D1} \bigg\},      
\end{eqnarray}      
\end{widetext}   
where   
$\hat{\mu}=\frac{1}{3}\mu_B  
+{\rm diag}_f(\frac{2}{3},-\frac{1}{3},-\frac{1}{3})\mu_Q  
+\lambda_3\mu_3+\lambda_8\mu_8$    
is the diagonal quark chemical potential matrix     
and $\hat{m}={\rm diag}_f(m_u,m_d,m_s)$ is    
the current-quark mass matrix.     
For $a=0$, $\tau_0=(2/3)^{1/2}{\mathbf 1}_f$, otherwise $\tau_a$ and      
$\lambda_a$  are Gell-Mann matrices acting in, respectively, flavor  
and color spaces.     
$C=i\gamma^2\gamma^0$ is the charge conjugation operator and      
$\bar{q}=q^\dagger\gamma^0$.     
The scalar quark-antiquark current-current interaction is given     
explicitely and has coupling strength $G_S$. The 3-momentum cutoff,    
$\Lambda$, is fixed by low-energy QCD phenomenology (see table I of    
\cite{Grigorian:2006qe}).    
The spin-0 and spin-1 diquark currents are    
$j_{D0,A}=q^TiC\gamma_5\tau_A\lambda_Aq$ and     
$j_{D1}=q^TiC(\gamma_1\lambda_7+\gamma_2\lambda_5+\gamma_3\lambda_2)q$.    
While the relative coupling strengths $\eta_{D0}$ and $\eta_{D1}$     
are essentially free parameters,  we restrict the discussion    
to the Fierz values, $\eta_{D0}=3/4$ and $\eta_{D1}=3/8$, see     
\cite{Buballa:2003qv}.   
Color superconducting phases in QCD with one flavor were first discussed    
in Refs. \cite{Schafer:2000tw,Alford:2002rz,Schmitt:2004et}, where it was   
also pointed out that the gap is of order 1 MeV   
in the spin-1 color-spin-locking (CSL) phase.   
This feature of the CSL phase is robust. See   
\cite{Aguilera:2005tg} for an analysis of our isotropic ansatz for the   
spin-1 diquark current, \cite{Aguilera:2006cj} for its generalization   
to the nonlocal case, and \cite{Marhauser:2006hy} for a self-consistent   
Dyson-Schwinger approach.   
   
\begin{figure}[h!tb]    
\includegraphics[angle=0,width=0.9\linewidth,clip=]{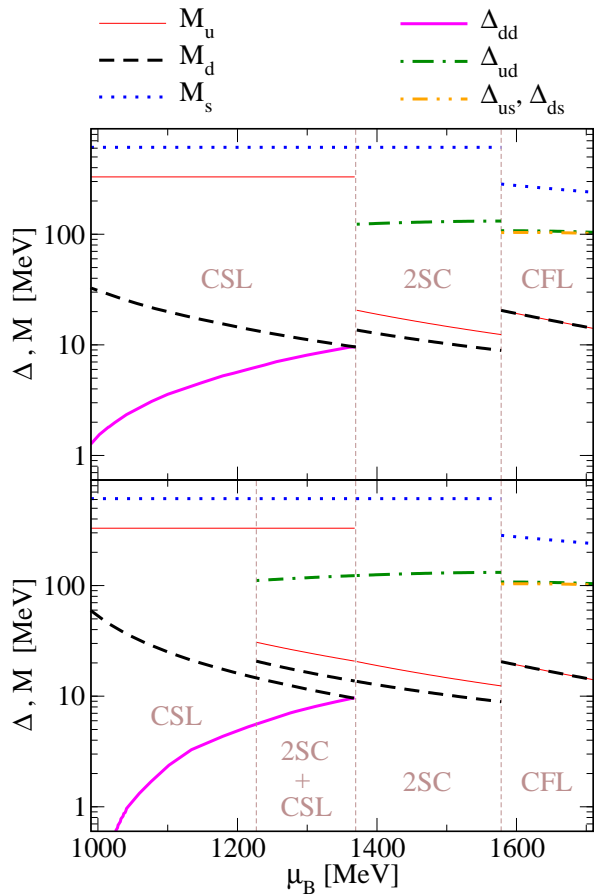}    
\caption{(Color online) Solution of the NJL gap equations for   
isospin-asymmetric charge-neutral matter. The upper (lower)   
panel corresponds to the hybrid EoS based on the DBHF (Shen)   
nuclear EoS. The asymmetry at a given value of the baryon   
chemical potential, $\mu_B$, is different in the   
two cases, because the charge density of nuclear matter depends   
on the model used.   
\label{f:gaps}}   
\end{figure}

\begin{figure}[htb]    
\includegraphics[angle=0,width=0.9\linewidth,clip=]{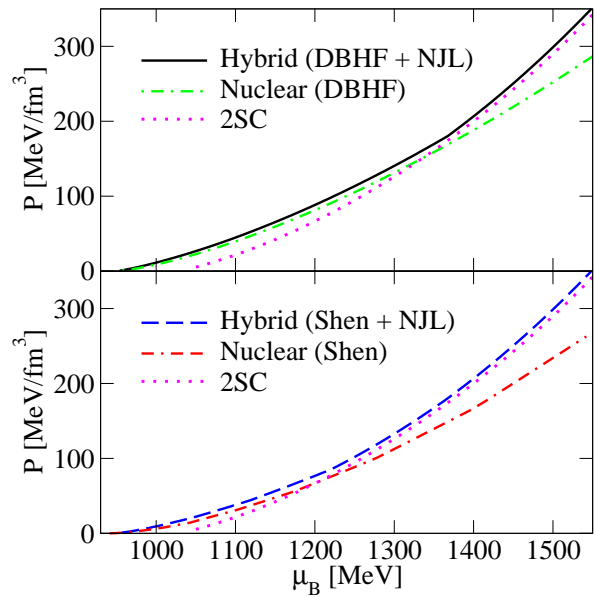}    
\caption{(Color online) Pressure of matter in beta-equilibrium for different   
nuclear-matter models 
{
(DBHF - upper panel, Shen - lower panel) and 
phase transition constructions with color superconducting quark matter 
(NJL model). 
Here DBHF (or Shen) + NJL refers to the mixed phase of nuclear matter 
and quark matter. 
At low chemical potential the quark-matter phase component is the negatively 
charged d-CSL phase, which lowers the asymmetry of the system and thereby 
gives a higher pressure of the mixed phase. 
At higher chemical potentials (1370 MeV for DBHF+NJL and 1230 for Shen+NJL) 
there is a transition in the quark sector to a 2SC phase component.}   
\label{f:pressure}}   
\end{figure}

The gaps and the renormalized masses are determined by minimization of the    
mean-field thermodynamic potential under the constraints of charge-neutrality  
and $\beta$-equilibrium, see Fig.~\ref{f:gaps}. For further details, see   
\cite{Blaschke:2005uj,Ruster:2005jc,Abuki:2005ms,Warringa:2005jh}.    
Different phases are characterized by different values of the   
order parameters (masses, gaps, etc.) and correspond to different   
local minima of the thermodynamic potential. For a particular   
choice of the baryon chemical potential there may be several local   
minima of the thermodynamic potential. The physical solution is   
that with lowest free energy, or, equivalently, highest pressure.   
In Fig.~\ref{f:pressure} we show the pressure of various phase   
constructions. 
Since we use separate models for the confined and deconfined states   
of quarks the dissociation of nucleons does not appear automatically   
within the model. Instead, for a given value of the baryon number  
chemical potential, three different phase constructions are considered:   
1) The homogenous charge-neutral and $\beta$-equilibrated nuclear   
matter phase. 2) The homogenous charge-neutral and $\beta$-equilibrated   
quark matter phase. 3) A charge-neutral equilibrium mixture   
of nuclear matter and quark matter, or two different quark-matter phases.   
For the models considered here we find that 
{
the asymmetry in three-flavor quark matter (CFL) is so small that it 
makes little sense to consider inhomogenous phase constructions.
For mixtures of nuclear matter with one-flavor (d-CSL) or two-flavor 
quark matter in the 2SC or normal quark matter (NQ) phase, however, 
the asymmetry is significantly lower and the pressure higher when compared 
with the homogeneous phases.}

\begin{figure}[htb]   
\includegraphics[angle=0,width=0.9\linewidth,clip=]{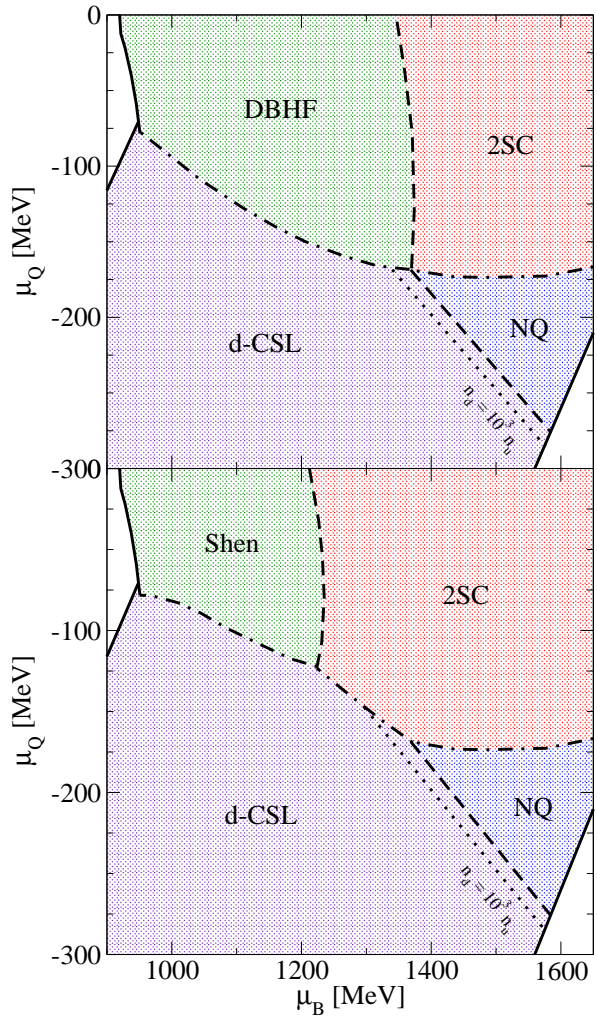}   
\caption{(Color online) Phase diagrams in the plane of baryon and charge  
chemical potential.  
The dash-dotted line denote the border between oppositely charged     
phases. The nuclear matter EoS is DBHF (upper) and Shen \etal (lower).   
Only one-flavor and two-flavor solutions are displayed, because three-flavor 
matter is charge-neutral for $\mu_Q \sim 0$. 
The transition to the nearly symmetric   
three-flavor CFL phase occurrs at $\mu_B \sim 1600$~MeV, see Fig.~\ref{f:gaps}.
{
At low densities the mixture of nuclear matter with one-flavor quark matter 
(d-CSL) is favored. At higher densities, beyond the up-quark threshold, 
a mixture with two-flavor quark matter is favored.
The two-flavor phases considered here are the normal quark
matter (NQ) and the superconducting (2SC) phase.} 
\label{f:phasediag}}    
\end{figure}

In Fig. \ref{f:phasediag} we plot the thermodynamically favored phase in   
the plane of baryon and charge chemical potentials.   
The hybrid EoS corresponds to the dash-dotted lines in Fig. \ref{f:phasediag} 
\ie the borders between positively and negatively charged phases, and they are 
constructed such that the {corresponding} mixture of nuclear matter, 
quark matter and leptons is charge neutral. 
{
At low densities a mixture with one-flavor quark matter is favored. 
At higher density, beyond the up-quark threshold, a mixture of one-flavor and
two-flavor quark matter is favored. 
The strange flavor occurs at still higher densities.
Note that the 2SC phase cannot persist at high $|\mu_Q|$ since the large 
difference in the Fermi levels of $u$ and $d$ quarks prevents their pairing
and the two-flavor quark matter is therefore in the normal phase (NQ).
} 

\begin{figure*}
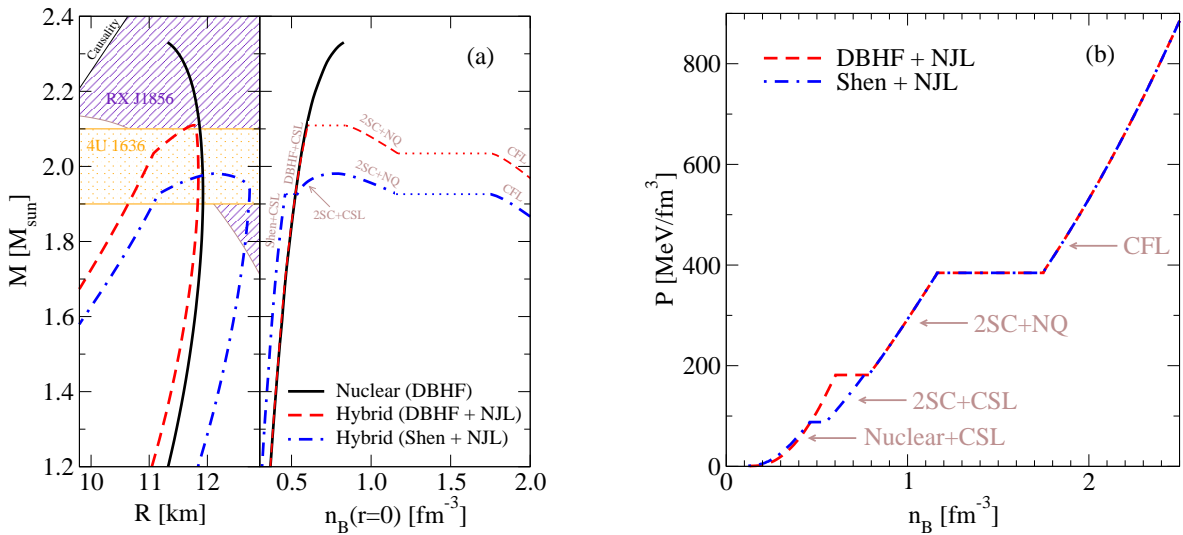
    
\begin{tabular}{ccc}    
\includegraphics[width=0.4\linewidth,clip=]{sequences2.eps} &   
\hspace{1cm} &    
\includegraphics[width=0.4\linewidth,clip=]{P_of_nB.eps}    
\end{tabular}    
\caption{(Color online)   
Compact star sequences (a) and hybrid equations of state (b).   
The phase structure at the center, $r=0$, changes with increasing density,   
as indicated in the figures. Constraints on the compact-star mass come   
from 4U 1636 \cite{Barret:2005wd} and on the mass-radius relation   
from RX J1856 \cite{Trumper:2003we}.   
\label{f:sequences}}   
\end{figure*}

Using the hybrid EoS we calculate compact-star   
sequences by solving the Oppenheimer-Volkoff equations for   
hydrostatic equilibrium. The hybrid-star sequences fulfill all modern   
constraints on the mass-radius relationship, see Fig. \ref{f:sequences}.     
For the DBHF hybrid EoS all stars with DBHF+CSL matter in the   
core are stable equilibrium solutions, while the appearance of   
u-quarks and the associated formation of a 2SC subphase renders   
the sequence unstable. The situation is somewhat different for   
the Shen hybrid EoS, because in addition to Shen+CSL stars there   
are stable solutions with 2SC+CSL matter in the core. In both   
cases configurations with strange quarks in the core are unstable.   
The hybrid star sequences 'masquerade' as neutron stars   
\cite{Alford:2004pf}, because the mechanical properties are   
similar to those of nuclear matter stars and the transition   
from nuclear matter to the mixed phase is associated with   
a relatively small discontinuity in the density.   
Unmasking neutron star interiors might therefore require   
observables based on transport properties, which could be   
strongly modified in presence of color superconductivity.    
It has been suggested to base such tests of the structure   
of matter at high density on an analysis of the cooling   
behavior 
\cite{Blaschke:2006gd,Popov:2004ey,Popov:2005xa,Grigorian:2006pu}    
or the stability of rapidly spinning stars against r-modes     
\cite{Madsen:1999ci,Drago:2007iy}.    
It has turned out that these phenomena are sensitive to   
the details of color superconductivity in quark matter.    
  
The down-quark chemical potential exceeds that of up quarks  
in asymmetric nuclear matter and, as we have illustrated above,  
this could lead to sequential deconfinement of the quark flavors.  
We have checked that the breaking of the $U(1)_A$ symmetry with  
a six-point 't~Hooft interaction does not rule out the single-flavor  
d-CSL solution, but the phase border is shifted to higher $|\mu_Q|$\ie  
more asymmetry is needed to realise the d-CSL phase in that case.  
A potential  
consequence of this is that the asymmetry of charge-neutral nuclear  
matter is less than that of d-CSL matter with broken $U(1)_A$ symmetry,  
and that the nuclear phase therefore is thermodynamically favored.  
As the origin of the $U(1)_A$ anomaly is unknown, see\eg the  
discussion in \cite{Blaschke:2005uj}, and the critical asymmetry  
depends on the parametrization of the NJL model and on the nuclear  
matter model used, a definite answer whether the d-CSL phase is  
realised is a matter of further investigation. Other effects of  
the inhomogenous phase mixture\eg Coulomb interactions and surface  
tension should also be considered in a future detailed investigation.  
Irrespective of these unsettled issues it is clear that the  
free energy of the d-CSL phase decreases with increasing asymmetry,  
in direct contrast to the behaviour of traditional phases such as  
the nuclear matter phase.  
In the following we discuss another interesting feature of the  
d-CSL phase, which could have important consequences for the  
phenomenology of compact stars. 
    
\section{Bulk viscosity and Urca emissivity of the single-flavor CSL    
phase}    
    
Rotating compact stars would be unstable against r-modes in the  
absence of viscosity \cite{Andersson:1997xt,Andersson:2000mf}.   
Constraints on the composition of compact-star interiors can   
therefore be obtained from observations of millisecond pulsars      
\cite{Madsen:1999ci,Drago:2007iy}.       
In such investigations the bulk viscosity is a key quantity and constraints 
on matter phases in neutron star interiors can be based on its value.   
Here we consider some relevant aspects of the bulk viscosity for color  
superconducting phases, starting with the 2SC phase and following   
the approach described in Ref. \cite{Sa'd:2006qv}.   
Note that the 2SC phase considered in \cite{Alford:2006gy}   
is a three-flavor phase, for which the nonleptonic process   
$u+d \leftrightarrow u+s$ is the dominant contribution.   
This process is not relevant for the 2SC phase discussed here, where  
the strange quark Fermi sea is not occupied.   
   
\begin{figure*}    
\begin{tabular}{ccc}   
\includegraphics[width=0.25\linewidth,clip=]{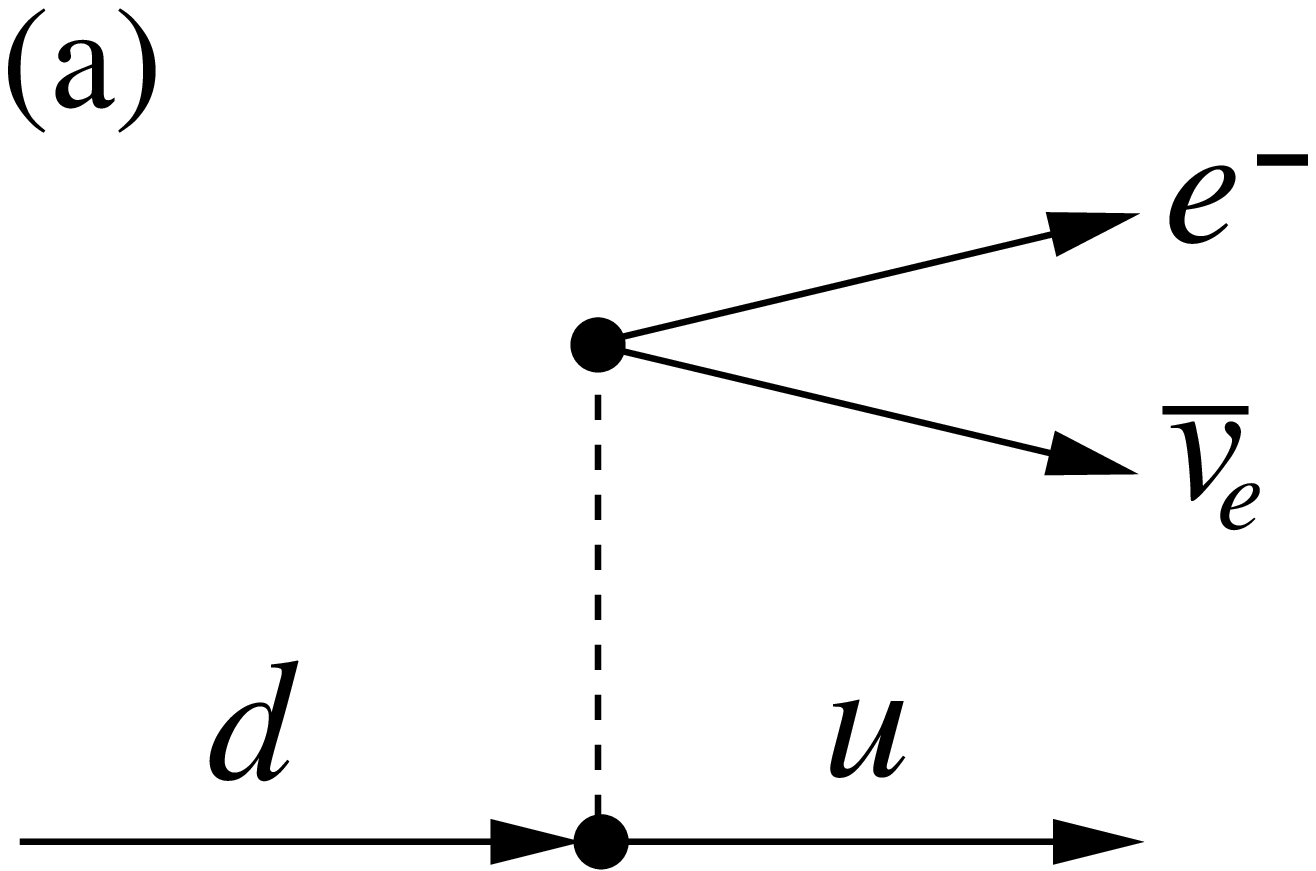}&   
\hspace{2cm} &    
\includegraphics[width=0.3\linewidth,clip=]{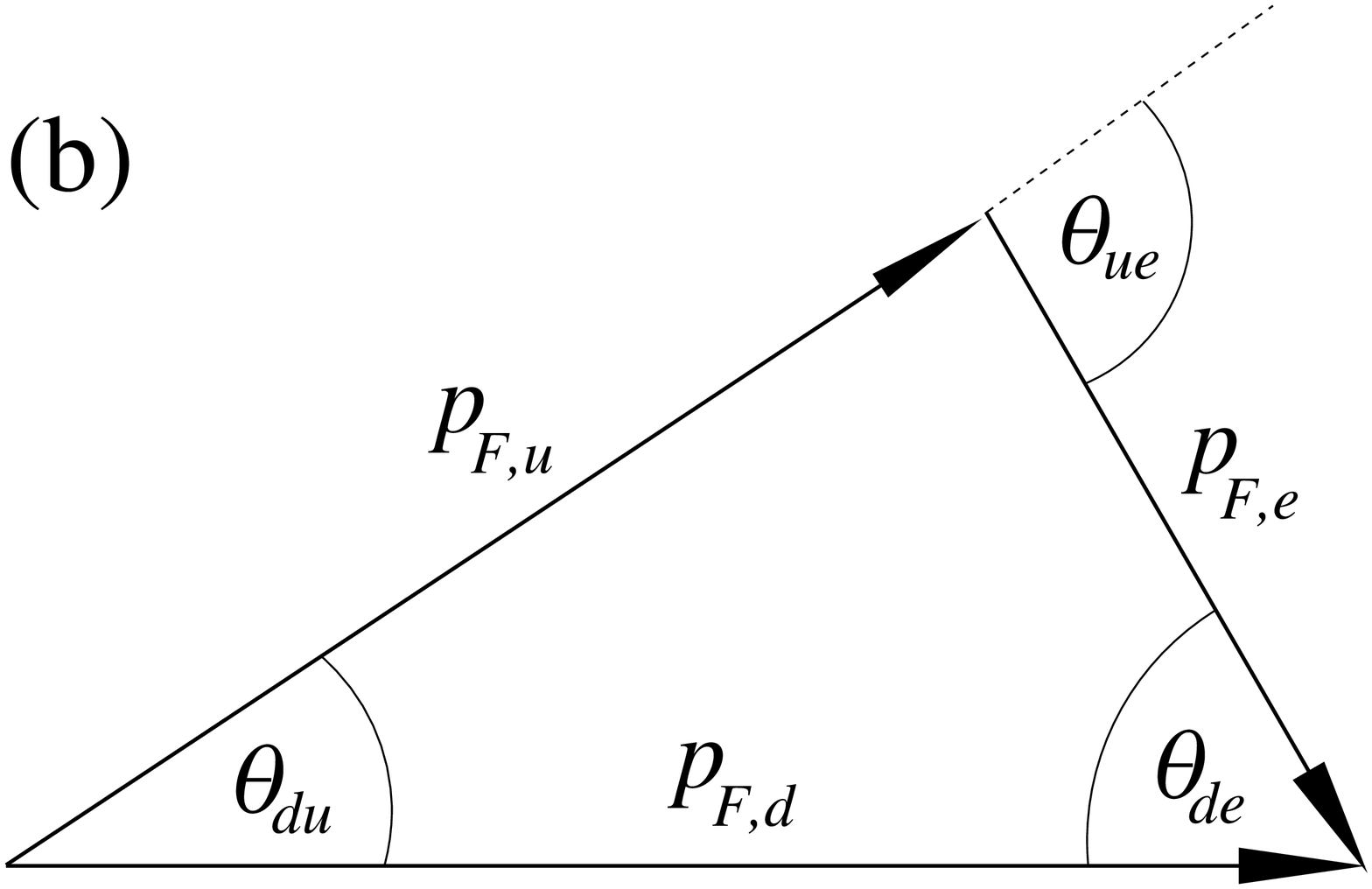}    
\end{tabular}   
\caption{(Color online)   
Direct Urca process in quark matter (a) and triangle   
of momentum conservation for it (b).   
\label{f:ptriangle}}   
\end{figure*}

The temperature-dependent bulk viscosity for the 2SC+CSL phase   
has been calculated self-consistently in \cite{Blaschke:2007bv}   
and is based on the flavor-changing weak processes of electron   
capture and $\beta$ decay    
\begin{eqnarray}    
u + e^- \rightarrow d + \nu_e~~,~~~d \rightarrow u + e^- + \bar{\nu}_e~.    
\end{eqnarray}     
It has been shown that the bulk viscosity is related to the direct   
URCA emissivity, which for normal quark matter was first calculated   
by Iwamoto \cite{Iwamoto:1982} and can be expressed as    
\begin{equation}    
\label{urca}    
\varepsilon_{0} \simeq     
\frac{914~\pi}{1680}G_F^2~\mu_e \mu_u \mu_d~ T^6~\theta_{ue}^2.     
\end{equation}    
Here $G_F$ is the weak coupling constant     
and $\theta_{ue}$ is the angle between the up-quark and electron     
momenta, which is obtained from momentum conservation in the matrix     
element, see Fig. \ref{f:ptriangle}.   
The triangle of momentum conservation holds for the late cooling stage,   
when the temperature is below 1 MeV and neutrinos are untrapped.    
Trigonometric relations are used to find an analytical expression   
for momentum conservation.   
To lowest order in $\theta_{de}$ the result is    
\begin{equation}    
\label{eq0}    
p_{F,d}-p_{F,u}-p_{F,e}\simeq - \frac{1}{2} p_{F,e}~\theta_{de}^2 ~.    
\end{equation}    
For small angles $\theta_{de} \simeq \theta_{ue}$, so it is possible to   
obtain an expression for the matrix element of the direct URCA process.    
Following Iwamoto \cite{Iwamoto:1982} one has to account either for    
quark-quark interactions to lowest order in the strong coupling   
constant, $\alpha_s$, (\ref{eq1}) or the effect of finite masses (\ref{eq2}):  
\begin{eqnarray}    
\mu_i &=& p_{F,i}\left(1+\frac{2}{3\pi}\alpha_s\right) ~,~~~{i=u,d}     
\label{eq1}\\    
\mu_i &\simeq&     
p_{F,i}\left[1+\frac{1}{2}\left(\frac{m_i}{p_{F,i}}\right)^2\right]~,     
~~{i=u,d,e}~~.     
\label{eq2}    
\end{eqnarray}    
From (\ref{eq0})-(\ref{eq2}) and the $\beta$-equilibrium condition,     
$\mu_d = \mu_u + \mu_e$, the angle $\theta_{de}$ that determines the   
emissivity (\ref{urca}) is obtained   
(\ref{eq0})    
\begin{eqnarray}\label{betaeq}    
\theta_{de}^2 \simeq \left\{ \begin{array}{cl}     
  \frac{4}{3\pi}\alpha_s \\[0.2cm]    
\frac{m_d^2}{p_{F,e}p_{F,d}}    
\left[1-\left(\frac{m_u}{m_d}\right)^2\left(\frac{p_{F,d}}{p_{F,u}}\right)    
-\left(\frac{m_e}{m_d}\right)^2\left(\frac{p_{F,d}}{p_{F,e}}\right)\right]    
\\    
\end{array}\right.. \nonumber\\   
\end{eqnarray}    
If interactions and masses are neglected, or the Fermi sea of one species is   
closed as in the single-flavor CSL phase, it follows that the triangle of   
momentum concervation in Fig. \ref{f:ptriangle} degenerates to a line or   
can not be closed.    
In that case the matrix element vanishes with the consequence that the direct  
URCA process does not occur, and also the bulk viscosity is zero.    
However, in the mixed nuclear+CSL phase there could be important friction   
and pair-breaking/formation processes, which we have not yet studied in detail.
This could be an interesting issue for further investigation due to the large  
difference in the masses of baryons and deconfined quarks.   

\section{Mechanism for deep crustal heating}    
   
Superbursts are rare, puzzling phenomena observed as a extremely long    
(4-14 hours) and energetic ($\sim ~10^{42} {\rm erg}$) type-I X-ray bursts    
from LMXBs.    
They take place if the accreted hydrogen and helium at the surface burns in    
an unstable manner, which is the normal case \cite{Stejner:2006tj}.      
As suggested in \cite{Page:2005ky} superbursts could originate from accreting  
strange stars with a thin crust and a core of three-flavor quark matter in   
the color-flavor-locked (CFL) phase.   
The suppression of the neutrino emissivity and heat conductivity in the   
CFL phase \cite{Blaschke:1999qx,Page:2000wt,Blaschke:2000dy}, caused by   
pairing gaps that affect all flavors, is of particular importance in this   
superburst scenario.   
Following Cumming et al. \cite{Cumming:2005kk} the underlying mechanism is 
unstable thermonuclear burning of carbon in   
the crust, at column depths of about $(0.5-3)\times 10^{12}~{\rm g~cm^{-2}}$.  
Carbon is a remnant of accreted hydrogen and helium at the surface.    
Observed superburst light curves suggest that the burning takes place at   
a depth where the crust reaches temperatures of about $6\times 10^8$ K   
and column depths of about $10^{12}~{\rm g~cm^{-2}}$.   
Such high temperatures in the crust at a certain depth are caused by    
deep crustal heating \cite{Haensel:1990,Ushomirsky:2001pd,Shternin:2007md}.  
The important ingredients for the strange-star model of superbursts  
\cite{Page:2005ky} are a thin baryonic crust of thickness 100 to 400~m,  
an energy release of 1 to 100 MeV per accreted nucleon by conversion into  
strange matter, a suppression of the fast direct URCA neutrino emissivity to  
the order of $10^{21}~{\rm~ erg~ cm^{-3}~ s^{-1}}$, and a thermal conductivity,
$\kappa$, of quark matter in the range  
$10^{19}-10^{22}~{\rm erg~cm^{-1}~s^{-1}~K^{-1}}$.  
      
\begin{figure}    
\includegraphics[angle=0,height=0.45\textwidth]{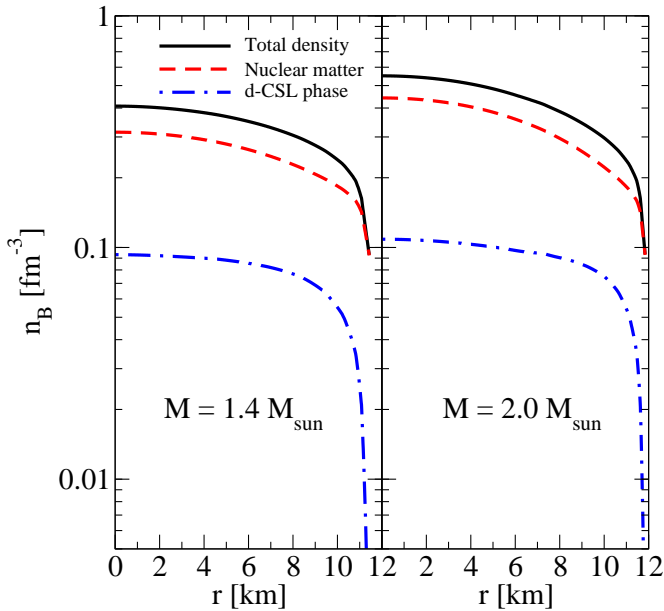}   
\caption{(Color online)    
Density profiles of two stars with masses $1.4$~M$_\odot$ and     
$2.0$~M$_\odot$.  
{In the model adopted here the mixed phase of d-CSL quark  
matter with nuclear matter extends up to the crust-core boundary.}  
\label{f:profile}}   
\end{figure}         
   
{
In Fig. \ref{f:profile} we show that d-CSL quark matter 
(in the mixed phase with nuclear matter) extends up to the crust-core boundary,
as strange quark matter does in the case of strange stars.
}
One of the main arguments for strange matter 
{
in the context of a 
superburst mechanism} is the fact that superconducting phases, like the CFL 
phase, can suppress fast neutrino emission processes of   
all quark flavors and are able to fulfill the fusion ignition condition.   
This is the case also for the single-flavor CSL phase. 
As we have shown above the fast direct URCA process is not possible at all in 
this phase, while slow neutrino cooling processes like bremsstrahlung of 
electrons and d-quarks exist.

We want to estimate the order of magnitude of the energy release $\Delta E$ 
due to partial conversion of ordinary nuclear matter to DBHF+CSL hybrid matter 
at the crust-core boundary.
As we apply the Gibbs construction of a phase transition a density dependent 
volume fraction of the d-quark admixture in the nuclear + CSL phase results, 
varying from zero to unity.
The caveat of this construction is that all thermodynamical quantities
at the onset of the phase transition vary continuously. 
However, in reality an infinitesimally small fraction of the d-quark subphase 
would imply that large residual color forces between d-quarks should occur.
Therefore, a solution of the phase admixture problem with a finite jump of the 
d-quark admixture at the onset of the tranistion should be energetically 
favored.
At the present stage of our work we cannot quantify this statement due to the 
absence of confining forces between color charges in our quark matter model.
An estimate which we would suggest here is to determine the fraction $\chi$ of 
dCSL matter at the d-quark dripline in the vicinity of the crust-core boundary.
Then one multiplies the change in energy per baryon due to the process 
$n \to ddu$ with $\chi$ as an estimate for the probablility of this process to 
occur per accreted nucleon.
A rough estimate (see Fig.~\ref{f:profile}) gives $0.001\le\chi\le 0.01$ which 
for a jump of the d-quark mass gap by $300$ MeV (see Fig.~\ref{f:gaps}) at the 
chiral transition (d-quark dripline) results in 
$0.6 \le \Delta E $ [MeV]$\le 6$ MeV.
This meets well the estimated range $\Delta E \sim 1 - 100$~MeV 
\cite{Page:2005ky,Cumming:2005kk} and could thus, in principle, 
explain burst ignition at appropriate depths for a suitable value of $\kappa$. 

Therefore, a strange matter core is not necessarily needed to resolve the  
superburst puzzle, because a hybrid-star model with quark matter in the d-CSL  
phase could have similar properties.  
Stejner \etal \cite{Stejner:2006tj} show that deep crustal heating    
mechanisms at the crust-core boundary\eg conversion of baryonic matter    
to strange quark matter, which can fulfill the constraints of the superburst   
scenario provide a consistent explanation also for the cooling of soft X-ray   
transients.    
Along the lines of this argument, we suggest that the d-quark drip effect,  
which leads to a mixture of nuclear matter with single-flavor quark matter  
in the CSL phase, can serve as a deep crustal heating mechanism. Superbursts  
and the cooling of X-ray transients are not only consistent with quark  
matter in compact stars but may qualify as a signature for its occurrence!

\section{Conclusions}    
In this paper we suggest a new quark-nuclear hybrid model for   
compact star applications that fulfills modern constraints from 
observations of compact stars.    
Due to isospin asymmetry, down-quarks may ``drip out'' from nucleons and     
form a single-flavor color superconducting (CSL) phase that is mixed     
with nuclear matter already at the crust-core boundary in compact stars.    
The CSL phase has interesting cooling and transport properties that are in    
accordance with constraints from the thermal and rotational evolution of     
compact stars \cite{Blaschke:2007bv}.     
It remains to be investigated whether this new compact star composition   
could lead to unambiguous observational consequences, 
and whether it is thermodynamically favored also when effects like  
Coulomb screening and surface tension are accounted for. 
We conjecture that the d-quark drip may serve as an effective deep crustal    
heating mechanism for the explanation of the puzzling superburst phenomenon    
and the cooling of X-ray transients.    
    
\section*{Acknowledgements}    
D.B. is supported in part by the Polish Ministry of Science and Higher     
Education under grants No. N N 202 0953 33, N N 202 2318 37 and by the Russian
Fund for Basic Research under grant No. 08-02-01003-a. 
T.K. is grateful for partial support from the Department of Energy, 
Office of Nuclear Physics, contract no.\ DE-AC02-06CH11357.    
The work of F.S. was supported by the Belgian fund for scientific research    
(FNRS).   
D.B. and F.S. acknowledge support from CompStar, a Research Networking 
Programme of the European Science Foundation.

\end{document}